%Paper: astro-ph/9206003
%From: HETV2@brownvm.brown.edu
%Date: Fri, 19 Jun 92 12:41:23 EDT

\input phyzzx
\baselineskip=24pt
\magnification=1000
\vsize=257 true mm   \advance\vsize by -1.9 true in
\hsize=200 true mm   \advance\hsize by -1.9 true in
\hoffset=.55 true in
\def\pp{\noindent\parshape 2 0truecm 15truecm 2truecm 13truecm}
\def\apjref#1;#2;#3;#4 {\ar\pp#1, {\it #2}, {\bf #3}, #4. \par}
\def\np{\vfill\eject}
\def\spose#1{\hbox to 0pt{#1\hss}}
\def\lta{\mathrel{\spose{\lower 3pt\hbox{$\sim$}}
        \raise 2.0pt\hbox{$<$}}}
\def\gta{\mathrel{\spose{\lower 3pt\hbox{$\sim$}}
        \raise 2.0pt\hbox{$>$}}}

\def\np{\vfill\eject}
\centerline{   }
\vskip 0.1in
\line{~~~~~~~~~~~~~~~~~~~~~~~~~~~~~~~~~~~~~~~~~~~~~~~~~~~~~~~~~
{}~~~~~~~~~~~~~~~~~BROWN-HET-765\hfill}
\line{~~~~~~~~~~~~~~~~~~~~~~~~~~~~~~~~~~~~~~~~~~~~~~~~~~~~~~~~~
{}~~~~~~~~~~~~~~~~~ June 1992\hfill}
\vskip 0.3in
\centerline{GAMMA RAY SIGNATURES FROM ORDINARY COSMIC STRINGS}
\vskip 0.2in
\centerline{Jane H. MacGibbon\footnote*{NAS/NRC Research Associate} }
\centerline{Code 665 NASA/Goddard Space Flight Center}
\centerline{Greenbelt, MD 20771}
\vskip 0.2in
\centerline{Robert H. Brandenberger}
\centerline{Brown University}
\centerline{Department of Physics}
\centerline{Providence, Rhode Island 02912}
\vskip 0.3in
\centerline{ABSTRACT}
\vskip 0.2in
We calculate the flux of ultra high energy photons from
individual ordinary (i.e. non-superconducting)
cosmic strings and compare the results with the sensitivity of
current and proposed TeV and EeV telescopes. Our calculations
give only upper limits for the gamma ray
flux, since the source of the photons, jets from
particle production at cusps, may be weakened by back reaction effects.
For the usual cosmic distribution of strings, the
predicted bursts from strings with the
value of mass per unit length associated with galaxy formation
or light strings may just be detectable. A
diffuse gamma ray background from light strings may
also be seen by the Fly's Eye detector at above $7\times 10^{10}$ GeV.

\np
\centerline{\bf I. INTRODUCTION}
\indent This work is motivated by the need for
direct evidence of cosmic strings$^1$. Even if
cosmic string theory succeeds in describing the large scale structure of the
universe$^2$, its validity will ultimately rest on the
detection of direct string signatures. Here we discuss the
chance of detecting ultra high energy photon radiation from the cusp
regions of individual nearby ordinary (i.e. non-superconducting)
cosmic strings.

\indent In a previous paper [Ref. 3; see also Refs. 4 \& 5], we investigated
the
possibility of detecting ultra high energy (UHE) neutrino radiation from the
cusps. We considered the case in which
the total energy in the cusp region is released as
extremely energetic particles almost instantaneously, about
once per oscillation of the string loop. These
primary emitted particles then decay down to some
superheavy fermion scale $Q_f$, at which point
we apply an extrapolation of the QCD multiplicity
function to determine the energy distribution of the final particles.
The value of $\mu$, the mass per unit string length,
required for galaxy and large-scale
structure formation is $G\mu/c^2\simeq 10^{-6}$.
We found that the UHE neutrino background from a distribution of strings with
this value lies below present observational bounds$^{6, 7}$
in the energy range $10^8$ GeV $<E< 10^{11}$ GeV, even if
the cusp mechanism is maximally effective. It is also smaller than
the photoproduced flux expected below
$E\simeq 10^{11}$ GeV from cosmic ray collisions with the
microwave background.  As $G \mu$ decreases, the cusp background
increases until $G\mu/c^2\simeq 10^{-15}$,
due to a greater number of small loops, and then decreases$^4$.
For $G \mu/c^2 \simeq 10^{-15}$ it may exceed the
observational bounds if $Q_f \gta 10^{15}$ GeV and
cusp annihilation works at maximal strength$^3$.
Cusp neutrinos may be more easily seen
above $E\simeq 10^{11}$ GeV; or by
detectors whose sensitivity matches
the expected photoproduced background around $10^{10}$ GeV (if
$10^{-15}\lta G\mu/c^2\lta 10^{-13}$ and $Q_f\gta 10^{15}$ GeV).
Note, however, that
the final energy distribution of decay neutrinos is
highly uncertain. At the energies we are concerned with,
the true neutrino flux may be either higher or lower than
our approximation by a couple of orders of magnitude.
In all cases, we found it extremely
unlikely that neutrino bursts from individual
cusps would ever be observed.

\indent The photon emission from cusp decays, integrated over the
strings in the Universe, should be very similar in shape and
magnitude to the neutrino emission. The neutrino
background $E^3 F (E)$ peaks at
about $E = 10^{-1} Q_f$ and then decreases as $E$ decreases.
The particles with lower energies today were emitted by strings
at higher redshifts. However, photons emitted with
energy $E\gta 10^6z^{-1}$ GeV
where $z$ is the redshift at the epoch of emission will have been affected
by pair production off the cosmic microwave
and radio backgrounds$^8$. Because of this, the
predicted photon background is less likely
to be observed than the neutrino background except above $10^{10}$ GeV.
On the other hand, the conclusions with regard to the
burst from an individual cusp
are reversed. Present TeV ($=10^3$ GeV) detectors
are much more sensitive to a
photon burst from strings because an
incoming photon has a much greater chance
of interacting with the Earth's atmosphere
than an incoming neutrino. In this paper,
we calculate the maximum photon cusp radiation from an individual string
or background of strings and compare
it with the detection capabilities of existing and proposed air
shower array and \v Cerenkov telescopes. Our main assumption is that
back reaction effects do not prevent cusps from forming, and that
once formed cusps copiously produce particles.
We conclude that searching for TeV
gamma-ray bursts or a $10^{11}-10^{12}$ GeV gamma-ray
background from strings probably represents the most likely way of
detecting cusp radiation from strings, if it occurs.

\indent Note that the recent detection$^{44}$ of anisotropies in the
microwave background by the COBE experiment is compatible$^{45}$ with cosmic
strings being the source of the seeds for structure formation and
$G\mu/c^2\simeq 10^{-6}$. However, many particle physics models predict strings
with a smaller mass per unit length. Such strings would be irrelevant for
structure formation, but might have other observable consequences. It is
important to look for independent constraints on such models. This is an
additional motivation for our work.

\indent In Section II, we recount cosmic string emission. In Section III we
discuss the probability of detecting local strings or a cosmic background of
strings in light of present and proposed TeV and EeV telescopes.
$c$ and $G$ represents the speed of light and Newton's
constant, respectively, $t_0$ denotes the present time and
$h$ is the Hubble parameter in units of $100$ kms$^{-1}$ Mpc$^{-1}$.
The scenario$^{9 - 11}$ in which superconducting cosmic strings$^{12,13}$
emit UHE cosmic rays once the current in the string reaches a critical
value, is unrelated to our process of cusp annihilation.

\centerline{\bf II. EMISSION FROM COSMIC STRINGS}

\indent Ordinary cosmic strings decay predominantly by
gravitational radiation, losing energy at a rate$^{14 - 16}$

$$ P_g =\gamma G\mu \mu c\eqno(2.1)$$
where $\mu$ is the mass per unit length
of the string and $\gamma$ is a constant of order $40-100$.
There are also two mechanisms for particle emission
from ordinary strings. Both are suppressed by a large factor with respect to
$P_g$, at all but the final stages of string life. In the
first mechanism, particle-antiparticle pairs are
produced in the background of a moving
string loop. Applying lowest order perturbation theory,
a string of microscopical width  $w \sim (\hbar/\mu c)^{1\over 2}$
radiates a power of$^{17}$

$$ P_a \sim {w\over R} \mu c^3\eqno(2.2)$$
$R$, the radius of the loop, is typically a cosmological length.

\indent The second mechanism, ``cusp annihilation''$^{18,19}$,
can be summarized as follows.
Ignoring the small but finite string width and describing the loop
trajectory by a world sheet $\underline{x}(s, \tau )$,
we can choose a gauge in Minkowski space for
which $\tau$ is the coordinate time $t$ and $s$ parametrizes the
length along the string.  The trajectories $\underline{x}(s,\tau )$
are then solutions of the equations which follow
from the Nambu action. These solutions are periodic
in time and typically contain one or more pairs of
cusps per oscillation - a cusp being a point $(s, t)$ on the string world
sheet where $| \underline{\dot x} | = 1$ and $\underline{x}^\prime = 0$
($^\prime$ denotes the derivative with respect to $s$).
At a cusp, the assumptions under which one
can show that string evolution
is described by the Nambu action break down.
Since two segments of the string overlap there, the
microphysical forces are very strong and should
lead to a smoothing out of the
cusp by particle emission.  Similar particle emission has
been shown$^{20}$ to occur close to the
interaction point between two long straight intercommuting
strings.  In this paper, we will assume that the entire
energy in the cusp region is released as particles. If we neglect
back reaction, this assumption seems reasonable.
However, back reaction may play a crucial role and prevent or diminish
the chance of cusps forming. (In which case, the
effect we are discussing would be much weaker.
We shall see, though, that a background from strings may
be detectable even if cusp annihilation does not work at full strength.)

\indent By expanding the solutions of the string equations of motion
about the cusp$^9$, one can show$^{18}$ that the s-parameter (comoving)
length of the region where the two string segments overlap is

$$\ell_c \sim w^{1/3} R^{2/3} \eqno(2.3)$$
The corresponding physical length obtained by evaluating
$\underline{x}(s, t)$ at $s = l_c$ is
$l_p\propto l_c^2 R^{-1}$. The energy per unit
comoving length is independent of the string velocity, whereas
the energy per unit physical length contains the usual relativistic
Lorentz factor, $\gamma_L= R/l_c$ (when evaluated
at $s=l_c$). Since the period of loop motion is $R/c$,
cusp annihilation produces a radiated power of

$$ P_c \sim {\mu\ell_c c^3\over R} \sim {\mu w^{1/3} c^3\over{R^{1/3}}}
\eqno(2.4)$$
averaged over the loop period. Each annihilation should
occur on the time scale associated with the energy scale of the
string$^{18}$, i.e. $\Delta t_{cusp}\simeq \ell_c/c$ in the frame comoving with
the loop, while the time scale between each cusp forming is
$R/c$. In the inertial frame, the initial particles produced at the cusp
will be beamed into a solid angle $\gamma_L^{-2}$.

\indent The primary particles emitted from the
cusps will be the scalar and gauge particles associated with the fields
which make up the cosmic string. These high energy particles then
decay rapidly into jets of lower mass products.
By conservation of quantum numbers,
cusp annihilations should produce
equal numbers of superheavy fermions and antifermions (up to CP
violation effects$^{21}$) which
decay into equal numbers of particles and
antiparticles (up to the initial charge
of the superheavy fermion and CP violation effects).
In order to calculate the photon flux, we need to know
how many photons of energy E are generated in the decay.
In the final decay stages, we assume that the fragmentation
proceeds via quarks, gluons and leptons and model it
in a way consistent with current QCD multiplicity data.
However, we should be cautious in extrapolating the empirical
QCD multiplicity functions to arbitrarily high energies: at energies above
the symmetry breaking scale of the field theory which gives
rise to strings

$$\sigma = (\mu\hbar c^3)^{1/2}\simeq 3.1\times 10^{19} \left(G\mu\over{c^2}
\right)^{1/2}\ \rm{GeV},\eqno(2.5)$$
a QCD-like extrapolation probably gives a poor description of fragmentation.
The uncertainty in the fragmentation pattern at energies above $\sigma$
introduces a large uncertainty in our calculations. We
could proceed thus in two ways.

\indent In the first approach, the approach taken in this paper,
we assume that the initial particles are emitted from the cusp with
energy $>> m_{pl}$ in the center of mass frame of the loop. The
particles then fragment after a number of steps into
particles with energy $Q_f << m_{pl}$, at which point we
apply the extrapolation of the QCD multiplicity
function for a jet of initial energy $Q_f$. We consider
various values of $Q_f$ around $\sigma$. For simplicity
the initial jet energy distribution is also assumed
to be monochromatic - extension to a more general
distribution is straightforward. The number of jets
with initial energy $Q_f$ emitted from a loop per cusp annihilation is thus

$$N ={\mu \ell_c c^2\over{Q_f}}\eqno(2.6)$$
Or per unit time and averaged over the period,

$$\dot N ={ P_c\over{Q_f}}\simeq {\mu^{5/6}\hbar^{1/6}c^{17/6}\over{
Q_fR^{1/3}}}\eqno(2.7)$$

\indent  In the second approach, the initial jet energy is
the energy of a Higgs particle emitted from the cusp in the center of mass
frame, i.e. $\sigma \gamma_L \propto \sigma {(\sigma R)}^{1/3}$.
We would then extrapolate the QCD fragmentation functions to that energy.
This has the effect of substantially increasing $Q_f$ in (3.13),
thereby decreasing the chances of detecting a cusp annihilation.
However, since it involves extending QCD-like behavior to energies above
the Planck scale, we regard it as less realistic and
do not consider it further here.

\indent The observable photon spectrum, in both approaches, is
dominated by the photons created by decaying neutral pions in
the quark and gluon jets$^{22}$. Any fragmentation function
should continue down to at least energies around the pion rest mass
$m_{\pi^0}\simeq 135$ MeV. Following Ref. 10 (and noting that QCD
jets generate roughly equal numbers of each charged
and neutral pion species), we describe
the pion multiplicity distribution in the jets by

$$dN^\prime/dx ={15\over{16}}x^{-3/2} (1-x)^2\eqno(2.8)$$
for simplicity, where $N^{\prime}(x)$ gives the
probability of finding a pion with energy $E_{\pi^0}= xQ_f$ in the jet.
Precise modelling of the multiplicity
function even up to initial jet energies
of $Q_f$ is not possible from first principles.
(A Drell-Yan-West approximation can be
applied as $x\longrightarrow 1$. Here however
we are mainly interested in the $x<<1$ region since
$Q_f$ is usually much greater than the energy thresholds of
detectors.) (2.8) implies that the final number
of pions in the jet scales as $\sqrt{Q_f}$. This
roughly equals the  multiplicity growth
seen in GeV-TeV collider experiments$^{10}$. The function also
matches well the more stringent $x<<1$ approximation derived in Ref. 23.
Since each pion decays into 2 photons,
the final photon distribution is found by
integrating (2.8) with invariant measure $dx/ x$
over all $E_{\pi^0}$ greater than the photon energy $E$.
Using (2.6), this gives a photon distribution of

$$ {{dN}\over {dE} } ={15\over{16}}{\mu \ell_c c^2\over{Q_f^2}}
\left( {16\over 3} -2x^{1/2} - 4x^{ -1/2} + {2\over 3} x^{-
3/2} \right) \Big|_{x = E/Q_f} \eqno(2.9)$$
All loops will be emitting photons
with energies between 0 and $Q_f$.

\indent In both approaches, the final decay products will be spread over a
solid angle $\Theta^2$ around the initial direction of the primary
particle. If $<p_T>$ is the average transverse momentum in a jet whose
initial energy is $Q$, $<N_{TOT}>$ is the mean total multiplicity and
$p_{TOT}$ is the total momentum ($p_{TOT}\simeq Q/c$ in the relativistic
limit), then$^{24}$ $\Theta\simeq <N_{TOT}><p_T>/p_{TOT}$. Using the results of
Ref. $24$ derived for QCD jets, $\Theta\simeq 1.29 \alpha_s(Q^2)$
radians to first order in the strong coupling ``constant'' $\alpha_s(Q^2)$
where

$$\alpha_s(Q^2)={12\pi\over{\left(33-2n_f\right)\ln\left(Q^2/\Lambda^2
\right)}}\left(1-{{6\left(153-19n_f\right)\over{\left(33-2n_f\right)^2}}
{\ln\left(\ln\left(Q^2/\Lambda^2\right)\right)\over{\ln\left(Q^2/\Lambda^2
\right)}}}\right)+....\eqno(2.10)$$
In (2.10), $n_f$ is the number of quark flavors with mass less than $Q$ and
$\Lambda$ is an empirically derived energy scale$^{25}$.
$\Lambda$ depends on $n_f$ in such a way that (2.10) remains valid for
all values of $Q$ at collider scales. Substituting $\Lambda^{\left(
n_f=4\right)}=238\pm 43$ MeV, we then have

$$\Theta\simeq{0.97\over{\ln\left(Q/\rm{GeV}\right)+1.44}}
\left(1-0.74{\ln\left(\ln\left(Q^2/5.7\times 10^{-2}\ \rm{GeV}^2\right)\right)
\over{\ln\left(Q^2/5.7\times 10^{-2}\ \rm{GeV}^2\right)}}\right)\ \rm{radians}
\eqno(2.11)$$
Again, the true form of $\Theta$ at initial jet energies $Q_f$ is
uncertain. However, because $\Theta^2>>\gamma_L^{-2}$, we can say that
{\it after decay} the emission from an annihilating cosmic string cusp
will be beamed into a solid angle $\Theta^2$, not $\gamma_L^{-2}$.
During the annihilation process, the direction of the beam will
continuously change at the rate given by $\gamma_L$. Following the method
of Rybicki and Lightman (Ref. $26$; see also Ref. $27$) and
noting that $\Delta t_{cusp}\simeq \gamma_L^{-1} R/c$, one can show
that an observer at Earth will remain in the cone of the beam defined by
$\Theta$ for a time

$$\Delta t_{beam}\simeq\Theta\gamma_L^{-1}\Delta t_{cusp}\eqno(2.12)$$
if $\gamma_L>>1$ and the timescale for the
decay of the emission is ignored. However,
$\Delta t_{beam}$ and $\gamma_L^{-1}\Delta t_{cusp}$ (the duration of cusp
annihilation in the inertial frame) are much smaller than the timescale
over which the emission decays. Thus
the duration of the signal at the detector should be determined by the
spreading out in arrival times of particles due to the decay process. Given
the uncertainties in the process, this is extremely difficult to estimate.
In the cascade process used to describe
QCD jet decay$^{28}$ prior to hadronization,
the lowest momentum non-relativistic decay products remain closest
to the creation point of the jet while the highest momentum ultra-
relativistic products travel farthest, reaching a distance $\hbar c/\Lambda$
in the center of mass frame from the creation point before hadronizing.
Thus an estimate of the spread in arrival times for particles created in a
QCD jet would $\hbar/\Lambda$ appropriately Lorentz transformed into the
observer's frame. If a similar analysis can be applied in the case of cusp
emission, an estimate of the duration of the burst observed at Earth could
be

$$\Delta t_{burst}\simeq {{\hbar}\over\Lambda}{{Q_f}\over\Lambda}\ .
\eqno(2.13)$$

\vskip 0.1in
\centerline{\bf III. EXPECTED SIGNAL AT DETECTOR}

\indent Let us now consider the number density of cosmic
strings given by the usual scale invariant
distribution.$^{1,29}$ We also include the effect of
gravitational and cusp radiation$^{18,19}$ on loops of small radius.
The number density $n(R)dR$ of loops at the present time $t_0$
with radius in the interval $[R, R +dR]$ is then given by$^{3,29}$

$$n (R) =\cases{\nu \left(ct_0\right)^{-2} R^{-2},
\ \ \ \ \ \ \ \ \ \ \ \ \ \ \ \ \ \ \ \ \ \rm{max}~(R_{eq},R_*)<R<ct_0\cr
\nu\left(ct_0\right)^{-2}R_{eq}^{1/2} R^{-5/2},
\ \ \ \ \ \ \ \ \ \ \ \ R_*<R<R_{eq},\ R_*<R_{eq}\cr
\nu \left(ct_0\right)^{-2}R_*^{-2},
\ \ \ \ \ \ \ \ \ \ \ \ \ \ \ \ \ \ \ \ \ R_{eq}<R<R_*,\ R_{eq}<R_*\cr
\nu\left(ct_0\right)^{-2}R_{eq}^{1/2} R_*^{-5/2},
\ \ \ \ \ \ \ \ \ \ \ \ R_{min}<R<\rm{min}~(R_{eq},R_*)\cr
0,\ \ \ \ \ \ \ \ \ \ \ \ \ \ \ \ \ \ \ \ \ \ \ \ \ \ \ \ \ \ \ \ \ \ \ \ \
R<R_{min}\cr}\eqno(3.1)$$
provided the following condition on $G\mu$ is met :
$t_0>5.4\times 10^{-44}\gamma^{-4}(G\mu)^{-9/2}$ sec, where

$$\eqalign{R_*&=\gamma G\mu t_0/c
\simeq 1.2\times 10^{30}\left(h\over{0.5}\right)^{-1}
\left(\gamma\over{10^2}\right)\left(G\mu\over{c^2}
\right)\ \rm{cm}\cr
R_{eq}&=ct_{eq}\simeq 5.8\times 10^{21} \left(h\over{0.5}\right)^{-4}\
\rm{cm}\cr
R_{min}&=ct_0^{3/4}(\sigma/\hbar)^{-1/4}\simeq
5.9\times 10^{12}\left(h\over{0.5}\right)^{-3/4}
\left(G\mu\over{c^2}\right)^{-1/8}\ \rm{cm}\cr}\eqno(3.2)$$
for an $\Omega=1$ Friedmann Universe.
Here $t_{eq}$ is the time of equal radiation and matter
density in the Universe, $h\simeq 0.3-1$ is the value of the
Hubble parameter today, $\gamma \simeq 40-100$ and $\nu$ is a constant
of order $0.01$ whose value must be determined in
numerical simulations.  Currently, $\nu$ is still uncertain by a
factor of at least 10 so we write

$$\nu = \nu_{0.01} 10^{-2}.\eqno(3.3)$$

\vskip 0.1in
\centerline{\bf A. BURST FROM INDIVIDUAL STRING}

\indent To estimate the probability of observing a burst
from an individual nearby string, we first note that
the observation of a cusp burst will be
characterized by the almost simultaneous arrival of
UHE photons from one position in the sky. Recall that
an estimate of the burst timescale at observation may be

$$\Delta t_{burst}\simeq {{\hbar}\over\Lambda}{{Q_f}\over\Lambda}
\simeq 10^{-8}\left({Q_f\over{10^{15}\ \rm{GeV}}}\right)
\ \rm{sec}\eqno(3.4)$$
arising from the jet decay process. The photons
will have the spectrum determined by the
decay of the initial superheavy fermions.
In the case of our approximation to the multiplicity
function (2.9), we would predict a slope of $E^{-3/2}$
at TeV energies ($<<Q_f$).

\indent Consider now a string loop of radius $R$
which is located a distance $d$ from Earth.
Noting that the radiation after decay will be beamed into
solid angle $\Theta^2$, the number per unit
energy per unit area of photons of energy $E$ expected at Earth
from a single cusp annihilation is

$$N_{burst}(E)\simeq {1\over{\Theta^2 d^2}} {\mu l_c c^2\over{Q_f^2}}
\left({16\over 3}-2\left({E\over{Q_f}}\right)^{1/2}-4\left({E\over{Q_f}}
\right)^{-1/2}+{2\over 3}\left({E\over{Q_f}}\right)^{-3/2}\right)
\eqno(3.5)$$
Here we have included the multiplicity approximation (2.9). The
predicted number of air showers at the detector above an
energy threshold $E_D$ is then

$$S_{burst}(>E_D)=\int_{E_D}^{\infty} N_{burst}(E) A_D\ dE\eqno(3.6)$$
where $A_D$ is the effective area of the detector.

\indent The telescope may also see the background of cosmic-ray induced
showers$^{30}$,

$$\dot{N}_{CR}(>E)\simeq 0.2\left(E\over{\rm{GeV}}\right)^{-1.5}\
\rm{cm}^{-2}\ \rm{sec}^{-1}\ \rm{sr}^{-1}\eqno(3.7)$$
This would produce a background in the detector of

$$S_{bgnd}(>E_D)\simeq 7\times 10^{-2}\Delta t_D\left(\theta\over{
10^{-3}\ \rm{sr}}\right)\left(A_D\over{10^{10}\ \rm{cm}^2}\right)
\left(E_D\over{10^5\ \rm{GeV}}\right)^{-1.5}\eqno(3.8)$$
showers per angular resolution $\theta$ of the detector
over the time $\Delta t_D>>\Delta t_{burst}$.
The new generation of atmospheric \v Cerenkov and air shower array
telescopes$^{31}$, currently under construction or proposed, typically
have energy thresholds of $E_D\simeq 10-100$ TeV,
collection areas in excess of $A_D\simeq 10^8-10^{10}$ cm$^2$
and angular resolutions less than $4\times 10^{-4}$ sr. They
should be able to reject hadron-induced showers from
photon-induced showers down to photon/hadron ratios of $10^{-4}-10^{-5}$.
For these specifications, the expected cosmic-ray or extragalactic photon
background in the telescope over the burst timescale $\Delta t_{burst}$
is negligible. Hence, to detect a cusp burst, we simply require that

$$S_{burst}(>E_D)\gta n_\gamma\eqno(3.9)$$
where $n_\gamma\simeq 1-5$ is the minimum number of showers
required to register as a burst.

\indent The logic of the burst analysis is as follows. In order for a burst
to be seen, the probability of a loop producing a cusp over a given period
of time must be sufficiently large. This leads to a condition $R<R_D$ for
loops to be observable (see below). Next, we note that the closest loop
with radius $R<R_D$ must lie within a distance $d_c(R,E_D)$ for its signal
to be strong enough in the detector. Thus, the mean seperation $d(R)$ of
loops of radius $R$ must be smaller than $d_c(R,E_D)$. As we shall see, for
$R<R_D$ the ratio $d_c(R,E_D)/d(R)$ decreases as $R$ decreases. Hence, the
condition which must be satisifed if we are to observe any burst is

$${d_c(R_D,E_D)\over{d(R_D)}}>1.$$

\indent Equations (2.3), (3.6)
and (3.9) imply that a string closer than

$$\eqalign{d_c(R,E_D)&\ {\eqalign {\simeq
\left(A_D\over{4\pi n_\gamma}\right)^{1/2}
&\Theta^{-1}\left({\mu l_c\over{Q_f}}\right)^{1/2} \left({E_D\over{Q_f}}
\right)^{1/2}\cr
&\left(-{16\over 3}+{4\over 3}\left({E_D\over{Q_f}}\right)^{1/2}
+8\left({E_D\over{Q_f}}\right)^{-1/2}
+{4\over 3}\left({E_D\over{Q_f}}\right)^{-3/2}\right)^{1/2}}}\cr
&\ {\eqalign{\simeq 1.0\times 10^9&\left(A_D/n_\gamma \over{\rm{cm}^2}
\right)^{1/2}\Theta^{-1}\left(\sigma\over{Q_f}\right)^{1/2}\left({\sigma
\over{10^{15}\ \rm{GeV}}}\right)^{1/3}\left({R\over{\rm{cm}}}\right)^{1/3}
\left({E_D\over{Q_f}}\right)^{1/2}\cr
&\cdot\left(-{16\over 3}+{4\over 3}\left({E_D\over{Q_f}}\right)^{1/2}
+8\left({E_D\over{Q_f}}\right)^{-1/2}+{4\over 3}\left({E_D\over{Q_f}}
\right)^{-3/2}\right)^{1/2}\ \rm{cm}\cr}}\cr}\eqno(3.10)$$
will be seen by the detector.
On the other hand, the average frequency of cusp
annihilations, $f(R)=cR^{-1}$.
Hence strings with radius $R_D\simeq 10^{18}-10^{19}$ cm should
produce cusps at a rate $0.1-1$ yr$^{-1}$ (which we regard as a
minimum detectable rate). For all values of $G\mu/c^2$,
$R_D$ is much less than $R_{eq}$ and much greater
than $R_{min}$. If

$${{G\mu}\over{c^2}} \gta 8.2\times 10^{-13}
\left(h\over{0.5}\right)\left(\gamma\over{10^2}\right)^{-1}
\left(R_D\over{10^{18}\ \rm{cm}}\right),
\eqno(3.11)$$
then $R_D$ is less than $R_*$. Because the chance of being in
the beam from an individual cusp is $\Theta^2/4\pi$, the closest observed
loop of radius less than or equal to $R_D$ should lie within a distance

$$\eqalign{d(R_D)&\simeq\left({\Theta^2\over 3}\int^{R_D}
_{R_{min}}n\left(R\right)dR\right)^{-1/3}\cr
&\simeq\cases{1.4\Theta^{-2/3}\nu^{-1/3}\left(ct_0\right)^{2/3}
R_{eq}^{-1/6}R_*^{5/6}R_D^{-1/3},
\ \ \ \ \ \ \ \ \ \ R_{min}<R_D<\rm{min}\ (R_{eq},R_*)\cr
1.4\Theta^{-2/3}\nu^{-1/3}\left(ct_0\right)^{2/3}R_{eq}^{-1/6}
\left({5\over 3} R_*^{-3/2} - {2\over 3}R_D^{-3/2}\right)^{-1/3},
\ \ \ \ \ R_*<R_D<R_{eq}\cr}}\eqno(3.12)$$
using (3.1). The ratio of the distances given in (3.10) and (3.12) becomes

$${d_c(R_D,E_D)\over{d(R_D)}}\simeq\cases{1\times 10^{-12}\left({A_D/
n_\gamma\over{10^{10}\ \rm{cm}^2}}\right)^{1/2}\Theta^{-1/3}\left({R_D
\over{10^{18}\ \rm{cm}}}\right)^{2/3}\left(\nu\over{10^{-2}}\right)
^{1/3}\left(h\over{0.5}\right)^{5/6}\cr
\ \ \ \ \ \ \cdot\left(\gamma\over{10^2}\right)^{-5/6}
\left(\sigma\over{10^{15}\ \rm{GeV}}\right)^{-5/6}
\left(Q_f\over{10^{15}\ \rm{GeV}}\right)^{-1}
\left(E_D\over{10^5\ \rm{GeV}}\right)^{1/2}\cr
\ \ \ \ \ \ \cdot\left(-{16\over 3}+{4\over
3}\left({E_D\over{Q_f}}\right)^{1/2}
+8\left({E_D\over{Q_f}}\right)^{-1/2}+{4\over 3}\left({E_D\over{Q_f}}
\right)^{-3/2}\right)^{1/2},\cr
\ \ \ \ \ \ \ \ \ \ \ \ \ \ \ \ \ \ \ \ \ \ \ \ \ \ \ \ \ \
R_{min}<R_D<\rm{min}
\ (R_{eq},R_*)\cr
\cr
1\times 10^{-11}\left({A_D/
n_\gamma\over{10^{10}\ \rm{cm}^2}}\right)^{1/2}\left({R_D
\over{10^{18}\ \rm{cm}}}\right)^{1/3}\left(\nu\over{10^{-2}}\right)
^{1/3}\left(h\over{0.5}\right)^{1/2}\cr
\ \ \ \ \ \ \cdot\left(\gamma\over{10^2}\right)^{-1/2}
\left(\sigma\over{10^{15}\ \rm{GeV}}\right)^{-1/6}
\left(Q_f\over{10^{15}\ \rm{GeV}}\right)^{-1}
\left(E_D\over{10^5\ \rm{GeV}}\right)^{1/2}\cr
\ \ \ \ \ \ \cdot\left(-{16\over 3}+{4\over
3}\left({E_D\over{Q_f}}\right)^{1/2}
+8\left({E_D\over{Q_f}}\right)^{-1/2}+{4\over 3}\left({E_D\over{Q_f}}
\right)^{-3/2}\right)^{1/2},\cr
\ \ \ \ \ \ \ \ \ \ \ \ \ \ \ \ \ \ \ \ \ \ \ \ \ \ \ \ \ \ R_*<R_D<R_{eq}\cr}
\eqno(3.13)$$
If we evaluate the fragmentation function
at $E_D=10$ TeV, the $(E_D/Q_f)^{-3/2}$ term dominates and so we finally have

$${d_c(R_D,E_D)\over{d(R_D)}}\simeq\cases{7\times 10^{-9}\left({A_D/n_\gamma
\over{10^{10}\ \rm{cm}^2}}\right)^{1/2}\Theta^{-1/3}\left({R_D
\over{10^{18}\ \rm{cm}}}\right)^{2/3}\left(\nu\over{10^{-2}}\right)
^{1/3}\left(h\over{0.5}\right)^{5/6}\cr
\ \ \ \ \ \ \cdot\left(\gamma\over{10^2}\right)^{-5/6}
\left(G\mu\over{c^2}\right)^{-5/12}
\left(Q_f\over{10^{15}\ \rm{GeV}}\right)^{-1/4}
\left(E_D\over{10^5\ \rm{GeV}}\right)^{-1/4},\cr
\ \ \ \ \ \ \ \ \ \ \ \ \ \ \ \ \ \ \ \ \ \ \ \ \ \ \ \ \ \ \ \ R_{min}<R_D<
\rm{min}\ (R_{eq},R_*)\cr
\cr
9\times 10^{-5}\left({A_D/n_\gamma
\over{10^{10}\ \rm{cm}^2}}\right)^{1/2}\Theta^{-1/3}\left({R_D
\over{10^{18}\ \rm{cm}}}\right)^{1/3}\left(\nu\over{10^{-2}}\right)
^{1/3}\left(h\over{0.5}\right)^{1/2}\cr
\ \ \ \ \ \ \cdot\left(\gamma\over{10^2}\right)^{-1/2}
\left(G\mu\over{c^2}\right)^{-1/12}
\left(Q_f\over{10^{15}\ \rm{GeV}}\right)^{-1/4}
\left(E_D\over{10^5\ \rm{GeV}}\right)^{-1/4},\cr
\ \ \ \ \ \ \ \ \ \ \ \ \ \ \ \ \ \ \ \ \ \ \ \ \ \ \ \ \ \ \ \ R_*<R_D<R_{eq}
\cr}\eqno(3.14)$$
At first glance this ratio lies well below $1$. This however
may not be so. There are large uncertainties in our knowledge
of $\nu$, $\gamma$ and the Hubble constant (which is probably $>0.5$).
More importantly, the true form of the fragmentation function
at these energies is unknown, as is the value of $Q_f$ at which we
can apply a QCD-like extrapolation. With regard to the
detector, the ratio can be increased by increasing the observing time
of the detector, thereby increasing $R_D$ in (3.14); by increasing
the effective area of the detector $A_D$; or by decreasing the
threshold energy $E_D$ of the detector. In summary, it is a
possibility that TeV detectors may register cusp bursts from an
individual cosmic string. We also stress
that the ratio (3.14) is derived assuming that the
cusp annihilation mechanism works at full efficiency.

\indent Since it may seem unnatural to fix $Q_f$, the mass of the
initial superheavy particles emitted
from the cusp, while varying $G\mu$,
we also set $Q_f = \sigma$ in which case (3.14) becomes

$${d_c(R_D,E_D)\over{d(R_D)}}\simeq\cases{6\times 10^{-10}\left({A_D
/n_\gamma\over{10^{10}\ \rm{cm}^2}}\right)^{1/2}\Theta^{-1/3}\left({R_D
\over{10^{18}\ \rm{cm}}}\right)^{2/3}\left(\nu\over{10^{-2}}\right)
^{1/3}\left(h\over{0.5}\right)^{5/6}\cr
\ \ \ \ \ \ \cdot\left(\gamma\over{10^2}\right)^{-5/6}
\left(G\mu\over{c^2}\right)^{-13/24}
\left(E_D\over{10^5\ \rm{GeV}}\right)^{-1/4},\cr
\ \ \ \ \ \ \ \ \ \ \ \ \ \ \ \ \ \ \ \ \ \ \ \ \ \ \ \ \ \ \ \ \ \ \ \ \ \
R_{min}<R_D<\rm{min}\ (R_{eq},R_*)\cr
\cr
7\times 10^{-6}\left({A_D
/n_\gamma\over{10^{10}\ \rm{cm}^2}}\right)^{1/2}\Theta^{-1/3}\left({R_D
\over{10^{18}\ \rm{cm}}}\right)^{1/3}\left(\nu\over{10^{-2}}\right)
^{1/3}\left(h\over{0.5}\right)^{1/2}\cr
\ \ \ \ \ \ \cdot\left(\gamma\over{10^2}\right)^{-1/2}
\left(G\mu\over{c^2}\right)^{-5/24}
\left(E_D\over{10^5\ \rm{GeV}}\right)^{-1/4},\cr
\ \ \ \ \ \ \ \ \ \ \ \ \ \ \ \ \ \ \ \ \ \ \ \ \ \ \ \ \ \ \ \ \ \ \ \ \ \
R_*<R_D<R_{eq}\cr}
\eqno(3.15)$$
If we chose $\gamma = 100$, $h = 0.75$, $\nu=0.03$ and
$G\mu/c^2 =3\times 10^{-7}$ (since these resemble the values used in the
cosmic string model of galaxy formation), the ratio becomes

$${d_c(R_D,E_D)\over{d(R_D)}}\simeq 4\times 10^{-6}\left({A_D
/n_\gamma\over{10^{10}\ \rm{cm}^2}}\right)^{1/2}\Theta^{-1/3}\left({R_D
\over{10^{18}\ \rm{cm}}}\right)^{2/3}
\left(E_D\over{10^5\ \rm{GeV}}\right)^{-1/4}\eqno(3.16)$$
Similarly for $G\mu/c^2\simeq 10^{-12}$, we have

$${d_c(R_D,E_D)\over{d(R_D)}}\simeq 2\times 10^{-3}\left({A_D
/n_\gamma\over{10^{10}\ \rm{cm}^2}}\right)^{1/2}\Theta^{-1/3}\left({R_D
\over{10^{18}\ \rm{cm}}}\right)^{1/3}
\left(E_D\over{10^5\ \rm{GeV}}\right)^{-1/4}\eqno(3.17)$$
for a typical TeV detector. Thus strings with small
$G\mu$ may be more easily seen.

\indent We must check that the cusp photons
are not cut off by pair-production off
the cosmic background photons in their travel to the detector.
The distance to the nearest string of radius $R_D$ should be

$$d(R_D)\simeq \cases{3\times 10^{10}\Theta^{-2/3}\left({R_D
\over{10^{18}\ \rm{cm}}}\right)^{-1/3}\left(\nu\over{0.03}\right)^{-1/3}
\left(h\over{0.5}\right)^{-5/6}\left(\gamma\over{10^2}\right)^{5/6}
\left(G\mu\over{c^2}\right)^{5/6}\ \rm{Mpc},\cr
\ \ \ \ \ \ \ \ \ \ \ \ \ \ \ \ \ \ \ \ \ \ \ \ \ \ \ \ \ \ \ \ \ \ \ \ \ \
R_{min}<R_D<\rm{min}\ (R_{eq},R_*)\cr
\cr
3\times 10^6\Theta^{-2/3}\left(\nu\over{0.03}\right)^{-1/3}
\left(h\over{0.5}\right)^{-1/2}\left(\gamma\over{10^2}\right)^{1/2}
\left(G\mu\over{c^2}\right)^{1/2}\ \rm{Mpc},\cr
\ \ \ \ \ \ \ \ \ \ \ \ \ \ \ \ \ \ \ \ \ \ \ \ \ \ \ \ \ \ \ \ \ \ \ \ \ \
R_*<R_D<R_{eq}\cr}\eqno(3.18)$$
from $(3.12)$. This compares to an absorption probability
of $\kappa_{\gamma\gamma}\simeq 10^{-3}$ Mpc$^{-1}$ for $E\simeq
10^5$ GeV photons in intergalactic space$^8$.
(The absorption rises quickly above $10^5$ GeV to
$\kappa_{\gamma\gamma}\lta 10^2$ Mpc$^{-1}$ for $E\simeq 10^6$ GeV
and then falls off with a slope of about $E^{-1}$. Below
$10^5$ GeV, the absorption rises to $\kappa_{\gamma\gamma}\lta
5\times 10^{-3}$ Mpc$^{-1}$ at $10^3-10^4$ GeV, due to pair-production
off light from Population II stars, and then falls off steeply.)
Thus the strings at a distance $d(R_D)$ should be just within the
visible Universe at $10^5$ GeV if $G\mu/c^2\simeq 3\times 10^{-
7}$. Strings with smaller values of $G\mu$ lie further within
the observable Universe at $10^5$ GeV.

\indent It is also relevant to mention the Fly's Eye detector$^{6,7}$.
The Fly's Eye detector is capable of seeing showers
induced by ultra-high energy photons,
although they have not yet been observed.
$10^9-10^{10}$ GeV photons should be visible out to $1-10$ Mpc,
while  $E\gta 10^{11}$ GeV photons will be affected by
pair-production off the Earth's magnetic field depending on their angle of
incidence$^{32}$. The effective area of the detector is $A_D\simeq 10^{13}$
cm$^2$ at $10^9$ GeV and increases slightly at higher energies.
Thus, if $G\mu/c^2\simeq 10^{-10}-10^{-11}$, (3.14) implies
that the Fly's Eye detector may offer a $2-3$ times greater chance
of detecting a burst at $10^9-10^{10}$ GeV than do TeV telescopes
at lower energies.

\vskip 0.1in
\centerline{\bf B. GAMMA RAY BACKGROUND FROM STRINGS}

\indent Regardless of whether individual bursts from cusps may be
detected, the combined radiation from cusps which have annihilated
over the history of the Universe will contribute a diffuse component
to the cosmic gamma ray background. To calculate the number density $F(E)dE$
of photons in the energy range $[E, E + dE]$, we must first
integrate over all times $t$ when photons with present energy $E$
were emitted.  At each $t$ we must also integrate over all loops
contributing to the emission. The number density of loops at an earlier
epoch is given by (3.1) and (3.2) with $t_0$ replaced by $t$,
for $R_{min}(t)\leq R_*(t)$ i.e. $t\geq t_B$ where

$$t_B= 5.4\times 10^{-44}\gamma^{-4}\left(G\mu/c^2\right)^
{-9/2}\ \rm{sec};$$
and

$$n (R) =\cases{\nu \left(ct\right)^{-2} R^{-2},
\ \ \ \ \ \ \ \ \ \ \ \ \ \ \ \ \ \ \ \ \ R_{eq}<R<ct\cr
\nu\left(ct\right)^{-2}R_{eq}^{1/2} R^{-5/2},
\ \ \ \ \ \ \ \ \ \ \ \ R_{min}<R<R_{eq}\cr}\eqno(3.19)$$
for $t\leq t_B$ sec. Thus using (2.7), we can write

$$F (E) = \int \ z^{-3} (t) f (z (t) E, t)\ dt\eqno(3.20)$$
where

$$f (z E, t) = {z\over{Q_f}} \> {dN\over{dx}}\Big|_{x=
zE/Q_f}\, \int_{R_{min}}^{ct} n (R, t) {{\mu^{5/6}\hbar^{1/6} c^{17/6} }
\over{Q_fR^{1/3}}}\ dR \eqno(3.21)$$
and the redshift $z$ has been included in the number density and energy.
($z=1$ in the present epoch.) It is also convenient to change
variables from $t$ to $z$. Hence we obtain

$$F (E) = {{\mu^{5/6}\hbar^{1/6} c^{17/6} }\over{Q_f^2}} \, \int dz \,
{dt\over{dz}} \, z^{-2} \, {dN\over{dx}}\Big|_{x=zE/Q_f}\,
\int^{t(z)}_0 \ {n (R,z)\over{R^{1/3}}}\ dR\eqno(3.22)$$
The $z$ integral runs over $[1, \rm{min}\left(Q_f/E,z_{CO}\right)]$
where $z_{CO}$ is the redshift at which a photon of energy
$z_{CO}E$ is cut off by interactions with the ambient matter or radiation
in the Universe. ($z_{CO}$ is discussed below.)

\indent Let us assume that $R_*(t)<R_{eq}$,
$zE<<Q_f$ (so that the $x^{-3/2}$ term in the $dN/dx$ approximation,
(2.9), dominates) and $z_{CO}<<z_{eq}$. The condition $t\leq,\geq t_B$
implies that three cases must be considered when integrating
(3.22): $z_{CO}\leq z_B$; $z_{CO}\geq z_B\geq 1$; and
$z_{CO}>1\geq z_B$ where $z_B=(t_0/t_B)^{2/3}$
$=5.2\times 10^{45}\left(\gamma/10^2
\right)^{8/3}\left(G\mu/c^2\right)^3$ is the redshift corresponding to
$t_B$.

\indent Firstly if $z_{CO}\leq z_B$, (3.22) becomes

$$\eqalign{E^3F(E)\simeq&{45\over {16}}\hbar^{1/6}c^{1/2}Q_f\left({E\over{Q_f}}
\right)^{3/2}\nu t_0\mu^{5/6}\left[ {10\over{11}}\left({\gamma G\mu
\over{c^2}}\right)^{-11/6}t_{eq}^{1/2}t_0^{-23/6}z^{3/4}\right.\cr
&\ \ \ \ -{1\over 4}
t_0^{-10/3}\ln z-{3\over{88}}t_{eq}^{-4/3}t_0^{-2} z^{-2}\cr
&\left.\ \ \ \ - {1\over 2}
\hbar^{1/12}c^{-1/4}\mu^{-1/12}\left({\gamma G\mu\over{c^2}}
\right)^{-5/2}t_{eq}^{1/2}t_0^{-4}z\right]^{z_{CO}(E)}_{z=1}\cr
\simeq &\left(E\over{0.1\ \rm{GeV}}\right)^{3/2}\left(Q_f\over{10^{15}\
\rm{GeV}}\right)^{-1/2}\left(\nu\over{10^{-2}}\right)\cr
&\ \ \ \ \cdot\left[ 2.9\times 10^{-10}
\left(\gamma\over{10^2}\right)^{-11/6}\left(G\mu\over{c^2}\right)^{-1}
\left(h\over{0.5}\right)^{5/6}z^{3/4}\right.\cr
&\ \ \ \ -2.0\times 10^4\left(G\mu\over{c^2}\right)^{5/6}
\left(h\over{0.5}\right)^{19/3} z^{-2}\cr
&\left.\ \ \ \ -5.3\times 10^{-22}
\left(\gamma\over{10^2}\right)^{-5/2}\left(G\mu\over{c^2}\right)^{-7/4}
\left(h\over{0.5}\right)z\right]^{z_{CO}(E)}_{z=1}
\ \rm{eV}^2\ \rm{m}^{-2}\ \rm{sec}^{-1}\cr}
\eqno(3.23)$$
If $z_{CO}\geq z_B \geq 1$,

$$\eqalign{E^3F(E)\simeq &{45\over {16}}\hbar^{1/6}c^{1/2}Q_f
\left({E\over{Q_f}}\right)^{3/2}\nu t_0\mu^{5/6}\Biggl\{ \left[ {10
\over{11}}\left(\gamma G\mu\over{c^2}
\right)^{-11/6}t_{eq}^{1/2}t_0^{-23/6}z^{3/4} \right.\cr
&\ \ \ \ -{1\over 4}t_0^{-10/3}
\ln z -{3\over{88}}t_{eq}^{-4/3}t_0^2 z^{-2}\cr
&\left.\ \ \ \ - {1\over 2}\hbar^{1/12}c^{-1/4}
\mu^{-1/12}\left(\gamma G\mu\over{c^2}\right)^{-5/2}t_{eq}^{1/2}t_0^{-4}
z\right]^{z_B}_{z=1}\cr
&\ \ \ \ +\left[{32\over{11}}\hbar^{-11/48}G^{-11/48}
c^{-55/48}\left(\gamma G\mu\over{c^2}
\right)^{11/48}t_{eq}^{1/2}t_0^{-27/8}z^{1/16}\right.\cr
&\left.\ \ \ \ -{1\over 4}t_0^{-10/3}\ln z -{3\over{88}}
t_{eq}^{-4/3}t_0^{-2} z^{-2}\right]^{z_{CO}(E)}_{z_B}\Biggr\} \cr}$$
$$\eqalign{\ \ \ \ \ \ \ \ \ \ \simeq &\left(E\over{0.1\ \rm{GeV}}\right)^{3/2}
\left(Q_f\over{10^{15}\ \rm{GeV}}\right)^{-1/2}
\left(\nu\over{10^{-2}}\right)\cr
&\ \ \ \ \cdot\Biggl\{\left[2.9\times 10^{-10}\left(\gamma
\over{10^2}\right)^{-11/6}\left(G\mu\over{c^2}\right)^{-1}
\left(h\over{0.5}\right)^{5/6}z^{3/4}\right.\cr
&\left.\ \ \ \ -5.3\times 10^{-22}
\left(\gamma\over{10^2}\right)^{-5/2}\left(G\mu\over{c^2}\right)^{-7/4}
\left(h\over{0.5}\right)z\right]^{z_B}_{z=1}\cr
&\ \ \ \
-\left[2.0\times 10^4\left(G\mu\over{c^2}\right)^{5/6}
\left(h\over{0.5}\right)^{19/3} z^{-2}
\right]^{z_{CO}(E)}_{z_B}\cr
&\ \ \ \ +
\left[8.9\times 10^{18}\left(G\mu\over{c^2}\right)^{17/16}
\left(h\over{0.5}\right)^{1/8}z^{1/16}
\right]^{z_{CO}(E)}_{z_B}\Biggr\}\ \rm{eV}^2\ \rm{m}^{-2}\ \rm{sec}^{-1}\cr}
\eqno(3.24)$$
and if $z_{CO}>1\geq z_B$,

$$\eqalign{E^3F(E)\simeq &{45\over
{16}}\hbar^{1/6}c^{1/2}Q_f\left({E\over{Q_f}}
\right)^{3/2}\nu t_0\mu^{5/6}\cr
&\ \ \ \ \cdot\left[{32\over{11}}\hbar^{-11/48}G^{-11/48}c^{-55/48}
\left(G\mu\over{c^2}\right)^{11/48} t_{eq}^{1/2}t_0^{-27/8}
z^{1/16}\right.\cr
&\left.\ \ \ \ -{1\over 4}t_0^{-10/3}\ln z
-{3\over{88}}t_{eq}^{-4/3}t_0^{-2} z^{-2}\right]^{z_{CO}(E)}_{z=1}\cr
\simeq &\left(E\over{0.1\ \rm{GeV}}\right)^{3/2}\left(Q_f\over{10^{15}\
\rm{GeV}}\right)^{-1/2}\left(\nu\over{10^{-2}}\right)\cr
&\ \ \ \ \left[8.9\times 10^{18}
\left(G\mu\over{c^2}\right)^{17/16}
\left(h\over{0.5}\right)^{1/8}z^{1/16}
\right]^{z_{CO}(E)}_{z=1}\ \rm{eV}^2
\ \rm{m}^{-2}\ \rm{sec}^{-1}\cr}
\eqno(3.25)$$

\indent The dominant interaction suffered by the extragalactic photons
is pair production off nuclei, if the photon energy at the relevant epoch
is $65$ MeV $\lta E'\lta 100$ GeV, or pair production off cosmic
background photons if $E'\gta 100$ GeV. The former
process cuts off the emitted photons at a redshift$^{33}$

$$z_{CO}\simeq 3.5\times 10^2 h^{-2/3}\left(\Omega_p\over {0.2}\right)
^{-2/3}\eqno(3.26)$$
where $\Omega_p$ is the present cosmological proton density
as a fraction of the critical density. Above $E'\simeq 100$ GeV,
we follow the method of Refs. 8 to calculate $z_{CO}$. The optical depth of
the universe to a photon emitted at a redshift $z'$ with
an energy corresponding to a redshifted energy today of $E$ is

$$\tau\left(E,z'\right)=\int^{z'}_1\kappa_{\gamma\gamma}\left(E,z\right)
{dl\over{dz}}\ dz\eqno(3.27)$$
Here $\kappa_{\gamma\gamma}\left(E,z\right)$
is the absorption probability per unit length and

$${dl\over{dz}}=cH_0^{-1}z^{-5/2}$$
for an $\Omega=1$ Friedmann universe, $H_0=100h$ km sec$^{-1}$
Mpc$^{-1}$ and $z'<z_{eq}$. Focussing on $z=1$ for the moment, we can write

$$\kappa_{\gamma\gamma}\left(E\right)={\pi e^4 m_e^2 c^4\over{E^2}}
\int^\infty_{m_e^2 c^4/E}\ \epsilon^{-2} n(\epsilon )\phi (\epsilon )\
d\epsilon\eqno(3.28)$$
where

$$\phi (\epsilon )=\int^{\epsilon E/\left(m_e^2 c^4\right)}_{1}\
{2m_e^2 c^4\sigma_{\gamma\gamma}\left(s\right)\over{\pi e^4}}\ ds$$
In (3.28), $\epsilon$ is the energy of the cosmic background photon,
$n(\epsilon)$ is the
number density per unit energy of cosmic background photons and
$\sigma_{\gamma\gamma}(s)$ is the total cross-section for the process
$\gamma +\gamma\longrightarrow e^+ + e^-$ as a function of the electron
or positron velocity in the centre of mass frame, $\beta =\left(
1- 1/s\right)^{1/2}$. The threshold for $e^+e^-$ pair production
is $\epsilon E=m_e^2 c^4$. We then find the cutoff redshift $z_{CO}(E)$ by
incorporating the relevant redshift dependence into $\epsilon$,
$n(\epsilon)$ and $\epsilon$ in $\phi(\epsilon)$ in (3.28) and
solving (3.27) for $\tau (E,z_{CO})=1$.

\indent The cosmic microwave background in the present era
extends between $2\times 10^{-6}$ eV $<\epsilon < 6\times 10^{-3}$ eV
and is accurately described$^{34}$ by $n(\epsilon)
=(\hbar c)^{-3} (\epsilon/\pi)^2$ $(\exp^{\epsilon/ kT} -1)^{-1}$
with $T_0=2.735(\pm 0.06)$ K. At earlier epochs,
$\epsilon$ should be replaced by
$z\epsilon$ and $T_0$ by $zT_0$.
A cosmic radio background has been observed$^{35}$
below $\epsilon\simeq 2\times 10^{-6}$ eV. The
$\epsilon n(\epsilon)$ radio spectrum
peaks at $\epsilon\simeq 10^{-8}$ eV
and extends down to $\epsilon\simeq 10^{-9}$ eV. At lower energies, the
cosmic background can not be seen because of inverse bremsstrahlung
(free-free) absorption of radio photons by electrons in the interstellar
medium. We will assume that the extragalactic radio spectrum continues
to fall off with the same slope down to $10^{-11}$ eV. The origin of the
radio background is not known but it is postulated to be the
integrated emission of all unresolved extragalactic radio sources$^{36}$
and to be modified below the peak by free-free
absorption by intergalactic gas$^8$. Since the evolution of
intergalactic gas is unknown, we will also assume for simplicity that
$\kappa_{\gamma\gamma}(E,z)\propto z$ at radio frequencies. The true
redshift dependence at these frequencies can effect our results little
since $z_{CO}\simeq 1$.

\indent Figure 1 plots $z_{CO}\left(E\right)$,
the cutoff redshift, as found from (3.27) using
a variation of Simpson's rule for numerical integration$^{37}$.
$z_{CO}$ decreases sharply to a minimum at $E\simeq 10^6$
GeV due to pair production off the microwave background and then falls off
gradually. Pair production off the radio background dominates at about
$5\times 10^9$ GeV with the greatest effect
at $E\simeq 4\times 10^{10}$ GeV. (The
maximum radio absorption and the energy at which it occurs are
two orders of magnitude smaller than the values presented
in Refs. 8. In those References, the radio background
was inexactly modelled prior to its observation.)
There is also an absorption component due to double pair
production$^{38}$ off the microwave background, $\gamma+\gamma\longrightarrow
e^++e^-+e^++e^-$. In the $s\longrightarrow\infty$ limit,
the cross-section for double pair production is
approximately constant, $\sigma'_{\gamma\gamma}\simeq 6.5\times
10^{-30}$ cm$^2$, and corresponds to an absorption probability of
$\kappa'_{\gamma\gamma}\simeq 6\times 10^{-27}$ cm$^{-1}$. This is
considerably weaker than the absorption probability for single pair
production if $10^5<E<10^{13}$ GeV. Applying the results of the
$z_{CO}$ calculation, we can restate in terms of $E$
the conditions given after (3.25) for
the dominant interaction: the photons are cut off
by pair production off nuclei
if their present energy would lie in the range $1.8\times
10^{-4}h^{2/3}\left(\Omega_p/0.2\right)^{2/3}$ $\lta E\lta$ $0.2-0.4$
GeV or by pair production off the microwave background
if $E\gta 0.2-0.4$ GeV.

\indent In Figure 2, we plot $E^3F(E)$ for various values of $G\mu/c^2$ and
the superheavy fermion scale $Q_f$. We can see from the curves that the
flux is greatest if $G\mu/c^2\simeq 10^{-15}$ and falls off
quickly for smaller values of $G\mu/c^2$ (due to the evaporation of
string loops by cusp and gravitational radiation). It also increases as $Q_f$
decreases. The $G\mu/c^2=10^{-15}$ flux is displayed in greater detail in
Figure 3. Since again it may be unnatural not to set $Q_f$ by the
symmetry breaking scale associated with the string, Figure 4 shows
$E^3F(E)$ for $Q_f=\sigma$. In this case, the flux is maximized at highest
energies if $G\mu/c^2\simeq 10^{-15}$.

\indent In all Figures, the dip at
$E\simeq 10^5$ GeV is produced by absorption off the microwave background.
This effect weakens above $10^6$ GeV.
Absorption off the radio background becomes important
at $E\simeq 10^{10}$ GeV, as marked by the kink in the spectra. Because of our
approximation to the multiplicity function (2.9), all spectra cut
off abruptly at $E=Q_f$. The true multiplicity function should
approach zero sharply$^{10}$ between $0.8Q_f<E<Q_f$. However, given the
uncertainity in extrapolating the collider multiplicity function to high
energies and the uncertainty in the cusp emission process and the initial
energy of the particles coming off the cusp (which we assumed to be
monochromatic for simplicity), further modelling of this
region is not justified.

\indent We also plot an
extrapolation of the observed $35-150$ MeV extragalactic gamma ray
data$^{39}$

$$E^3F(E)=3.3\left(\pm 0.6\right)\times 10^{16}\left(E\over{0.1\ \rm{GeV}}
\right)^{0.6\pm 0.2}\ \rm{m}^{-2}\ \rm{sec}^{-1}\ \rm{eV}^2\eqno(3.29)$$
on Figures 2-4. No measurements of the diffuse gamma ray background
above $150$ MeV have been made. The EGRET experiment, currently flying, will
be capable of detecting photons up to $20$ GeV.
It is not known, though, if the extragalactic background extends
to energies above $150$ MeV or, if it exists, if it
can be resolved - even at high Galactic latitudes - out of the Galactic
background which falls off less steeply
around $100$ MeV. Between $10^4-10^6$ GeV,
there is also a competing predicted
diffuse flux arising from the interaction of
extragalactic UHE cosmic rays with the microwave background$^{40}$. In this
process, cosmic rays pair-produce and inverse Compton scatter off the
microwave background creating cascade photons.
The postulated cosmic ray-induced flux is
of order $E^3F(E)\simeq 10^{20}$ eV$^2$ m$^{-2}$ sec$^{-1}$ at its peak
($E\simeq 10^5$ GeV) and falls off by more than 3 orders of
magnitude between $10^5$ and $3\times 10^5$ GeV.
Thus even for the lowest values of $Q_f$ and $G\mu/c^2
\simeq 10^{-15}$, emission from cosmic string cusp annihilation is
unlikely to be detected between $10^4-10^6$ GeV.

\indent We conclude that
the most sensitive regime to search for a string background is
above $E\simeq 10^{11}$ GeV. At these energies, unlike TeV energies, air
shower detectors can not distinguish between photon-induced and cosmic
ray-induced showers. However, the diffuse cosmic-ray background is expected
to be cut off above the Greisen energy ($E\simeq 7\times 10^{10}$ GeV) by
pair production of charged pions off the microwave background$^{41}$.
The $10^8-10^{10}$ GeV
cosmic ray data are consistent with this prediction. The Fly's Eye cosmic
ray measurements$^6$ are shown on Figures 2-4 and lie considerably above the
string-generated flux for $10^8<E<7\times 10^{10}$ GeV. Above
$E\simeq 7\times 10^{10}$ GeV, on the other hand,
where there should be no competing
background, a string signature should stand out provided it is
greater than the minimum flux which can be detected by the telescope. If the
telescope has an effective area $A_D$, the number of showers seen by the
detector in an observing time $\Delta t_D$ is

$$S(>E)\simeq 0.1\left({E^3F(E)\over{10^{21}\ \rm{eV}^2\rm{m}^{-2}
\rm{sec}^{-1}}}\right)\left(E\over{10^{11}\ \rm{GeV}}\right)^{-2}
\left(A_D\over{10^{14}\ \rm{cm}^2}\right)\left(\Delta t_D\over
{1\ \rm{yr}}\right)\eqno(3.30)$$
(To derive (3.30), note from the Figures
that approximately $F(E)\propto E^{-1.25}$
above $E\simeq 10^{10}$ GeV). The current configuration of
the Fly's Eye telescope has an effective aperature of $10^{13}$ cm$^2$ sr at
$E\simeq 10^{11}$ GeV; the High Resolution Fly's Eye detector$^{42}$
(HiRes) presently under construction will
have an effective aperature of $7\times 10^{13}$
cm$^2$ sr at $E\gta 10^{11}$ GeV. We also note that
$E\gta 10^{11}$ GeV photons arriving perpendicular to the Earth's magnetic
field will be affected by pair-production off the magnetic field$^{32}$.
Thus, returning to the Figures,
a cosmic string background would be detectable at
$E\simeq 10^{12}$ GeV over $\Delta t_D=1$ yr
if, for example, $10^{12}\lta Q_f\lta 10^{14}$ GeV and $G\mu/c^2
\simeq 10^{-13}$, $10^{12}\lta Q_f\lta 10^{18}$ GeV and $G\mu/c^2
\simeq 10^{-15}$ (see Figures 1 and 3) or if $Q_f\simeq\sigma$
and $G\mu/c^2\simeq 10^{-15}-10^{-13}$ (see Figure 2).
We stress that the predicted flux from cusp annihilation is
uncertain, particularly at these energies, due to the inexact knowledge of
the cusp annihilation process and the extrapolation of particle decay to
ultra high energies. The true flux may be greater or less than
shown in the Figures and may have a different spectral index. It
is of note, however, that a string background may be
detectable even if the annihilation process does not work at full
efficiency. For example, taking our approximation in Figure 2, even if
the process worked at $1\%$ efficiency (i.e. $1\%$ of the energy in the
cusp region is converted into particles), strings with $G\mu/c^2
\simeq 10^{-15}$ and $Q_f\simeq\sigma$ would produce one observable shower
per year.

\vskip 0.1in
\centerline{\bf IV. CONCLUSIONS}
\indent In the usual cosmic string scenario
of galaxy formation with $G\mu /c^2\simeq 10^{-6}$,
it may be just possible to detect
ultra-high energy gamma-ray bursts from the
cusp annihilation of nearby strings, if such radiation occurs.
If $G\mu$ is lower, the emission may be more easily seen.
Because the probability of detecting the bursts
is still small, we cannot yet derive new
lower bounds on $G\mu$, which would complement
the upper bounds found by considering the
distortion of binary star systems in
a string-produced background of gravitational radiation$^{43}$
and the COBE results for the quadrupole anisotropy of the microwave
background$^{45}$.

\indent If $G\mu /c^2\simeq 10^{-15}-10^{-13}$, the diffuse gamma-ray
background from cusp emission may be detected also by EeV telescopes such
as the Fly's Eye and Fly's Eye HiRes experiments.
This signal may be seen even if the cusp evaporation
process does not work at full efficiency.

\vskip 0.2 in
\centerline{\bf ACKNOWLEDGEMENTS}
\indent We thank Floyd Stecker,
Michael Salamon, Rob Preece, Terry Walker
and Michael Turner for useful conversations.
This work was supported in part (at Brown) by the
U.S. Department of Energy under grant No.
DE-AC02-76ERO3130 Tasks K $\&$ A and by the
Alfred P. Sloan Foundation.

\bigskip
\centerline{\bf REFERENCES}
\item{1.~} For reviews on cosmic strings, see e.g.
T.W.B. Kibble, {\it Phys. Rep.} {\bf 67}, 183 (1980);\hfill \break
A. Vilenkin, {\it Phys. Rep.} {\bf 121}, 263 (1985).
\item{2.~} For recent reviews see e.g.
R. Brandenberger, {\it Int. J. Mod. Phys.} {\bf A2}, 77 (1987); \hfill \break
N. Turok, ``Phase transitions as the origin of large scale structure in the
universe'', in {\it Particles, Strings and Supernovae}, Proceedings of
TASI 1988, Brown Univ., ed. by A. Jevicki and C.-I. Tan
(World Scientific, Singapore, 1989).
\item{3.~} J. MacGibbon and R. Brandenberger, {\it Nucl. Phys.} {\bf B331},
153 (1990).
\item{4.~} P. Bhattacharjee, {\it Phys. Rev.} {\bf D40}, 3968 (1989).
\item{5.~} P. Bhattacharjee and N. Rana, ``Ultrahigh-energy particle flux from
cosmic strings'', Fermilab preprint (1990).
\item{6.~} R. Baltrusaitis et al., {\it Astrophys. J.}, {\bf 281} L9
(1984);  R. Baltrusaitis et al., {\it Phys. Rev. Lett.}
{\bf 54}, 1875 (1985).
\item{7.~} R. Baltrusaitis et al., {\it Phys. Rev.} {\bf D31}, 2192
(1985).
\item{8.~} R. Gould and G. Schreder, {\it Phys. Rev.} {\bf 155}, 1404 \& 1408
(1967); \hfill \break
F. Stecker, ``Cosmic gamma rays'', NASA publication SP-249 (1971).
\item{9.~} D. Spergel, T. Piran and J. Goodman, {\it Nucl. Phys.}
{\bf B291}, 847 (1987).
\item{10.~} C. Hill, D. Schramm and T. Walker, {\it Phys. Rev.}
{\bf D36}, 1007 (1987).
\item{11.~} V. Berezinskii and H. Rubinstein, {\it Nucl. Phys.} {\bf B323}, 95
(1989).
\item{12.~} J. Ostriker, C. Thompson and E. Witten, {\it Phys. Lett.}
{\bf 180}B, 231 (1986).
\item{13.~} E. Witten, {\it Nucl. Phys.} {\bf B249}, 557 (1985).
\item{14.~} A. Vilenkin, {\it Phys. Rev.} {\bf D23}, 852 (1981); {\it
Phys. Lett.} {\bf 107B}, 47 (1982).
\item{15.~} T. Vachaspati and A. Vilenkin, {\it Phys. Rev.} {\bf D31},
3052 (1985).
\item{16.~} N. Turok, {\it Nucl. Phys.}, {\bf B242}, 520 (1984).
\item{17.~} M. Srednicki and S. Theisen, {\it Phys. Lett.} {\bf 189B},
397 (1987).
\item{18.~} R. Brandenberger, {\it Nucl. Phys.} {\bf B293}, 812
(1987).
\item{19.~} R. Brandenberger and A. Matheson, {\it Mod. Phys. Lett.}
{\bf A2}, 461 (1987).
\item{20.~} E.P.S. Shellard, {\it Nucl. Phys.} {\bf B283}, 624
(1987);\hfill \break
E.P.S. Shellard and P. Ruback, {\it Phys. Lett.} {\bf 209B}, 262
(1988);\hfill \break
K. Moriarty, E. Myers and C. Rebbi, {\it Phys. Lett.} {\bf 207B}, 411
(1988);\hfill \break
R. Matzner, Computers in Physics, Sept./Oct. 1988, 51.
\item{21.~} M. Kawasaki and K. Maeda, {\it Phys. Lett.} {\bf B209},
271 (1988);\hfill \break
R.H. Brandenberger, A.C. Davis and M. Hindmarsh, {\it Phys. Lett} {\bf
B263}, 239 (1991).
\item{22.~} W. Bartel et al, {\it Z. Phys.} {\bf C28}, 343 (1985); \hfill
\break
D. Scharre et al, {\it Phys. Rev. Lett.} {\bf 41}, 1005 (1978).
\item{23.~} C. Hill, {\it Nucl. Phys.} {\bf B224}, 469 (1983).
\item{24.~} P.E.L. Rakow and B.R. Webber, {\it Nucl. Phys.} {\bf B191}, 63
(1981).
\item{25.~} M. Agiular-Benitez et al, {\it Phys. Letts.} {\bf B239} (1990).
\item{26.~} G. Rybicki and A. Lightman, {\it Radiative Processes in
Astrophysics} p. 170(New York, Wiley, 1979).
\item{27.~} A. Babul, B. Paczynski and D. Spergel, {\it Ap. J.} {\bf 316},
L49(1987); \hfill \break
B. Paczynski, {\it Ap. J.} {\bf 335}, 525 (1988).
\item{28.~} G. Marchesini and B.R. Webber, {\it Nucl. Phys.} {\bf B238},
1 (1984).
\item{29.~} A. Albrecht and N. Turok, {\it Phys. Rev. Lett.} {\bf 54},
1868 (1985); \hfill \break
D. Bennett and F. Bouchet, {\it Phys. Rev. Lett.} {\bf 60}, 257 (1988); \hfill
\break
B. Allen and E.P.S. Shellard, {\it Phys. Rev. Lett.} {\bf 64}, 119 (1990).
\item{30.~} F. Halzen, E. Zas, J. MacGibbon and T. Weekes,
{\it Nature} {\bf 353}, 807 (1991).
\item{31.~} C. Akerlof et al, {\it Nucl. Phys. B (Proc. Suppl.)} {\bf 14A},
237 (1990).
\item{32.~} P.A. Shurrock {\it Ap.J.} {\bf 164}, 529 (1971).
\item{33.~} J.H. MacGibbon and B.J. Carr, {\it Ap. J.} {\bf 371}, 453
(1991).
\item{34.~} J. Mather, {\it Ap. J. Letts.} {\bf 354}, L37(1990).
\item{35.~} T.A. Clark, L.W.Brown and J.K. Alexander, {\it Nature} {\bf 228},
847(1970);\hfill \break
A.H. Bridle, {\it M.N.R.A.S.} {\bf 136}, 14(1967).
\item{36.~} M.T. Ressell and M.S. Turner, {\it Comments on Astrop.} {\bf 14},
323(1990).
\item{37.~} J.B. Scarborough, {\it Numerical Integration} Oxford University
Press, p. 136(1962).
\item{38.~} R.W. Brown, W.F. Hunt and K.O. Mikaelian, {\it Phys. Rev.}
{\bf D8}, 3083(1973).
\item{39.~} C.E. Fichtel et al, {\it Ap. J.} {\bf 222}, 833
(1978).
\item{40.~} F. Halzen et. al., {\it Phys. Rev.} {\bf D41}, 342(1990).
\item{41.~} K. Greisen, {\it Phys. Rev. Lett.} {\bf 16},  748(1966);
\hfill\break
C.T. Hill and D.N. Schramm, {\it Phys. Rev.} {\bf D31}, 564(1983);
\hfill\break
C.T. Hill and D.N. Schramm, {\it Phys. Letts.} {\bf 131B},
247(1983).
\item{42.~} L. Borodovsky et al, {\it Proc. 22nd Intl. Cosmic Ray Conf.}
Dublin Ireland {\bf 2}, 688 (1991)\hfill\break
W.Au et al, {\it Proc. 22nd Intl. Cosmic Ray Conf.}
Dublin Ireland {\bf 2}, 692 (1991).
\item{43.~} C. Hogan and M. Rees, {\it Nature} {\bf 311}, 109 (1984);\hfill
\break
E. Witten, {\it Phys.Rev.} {\bf D39}, 272 (1984);\hfill \break
R. Brandenberger, A. Albrecht and N. Turok, {\it Nucl.Phys.} {\bf B277},
605 (1986);\hfill \break
F. Accetta and L. Krauss, {\it Nucl. Phys.} {\bf B319}, 747 (1989);\hfill
\break
N. Sanchez and M. Signore, {\it Phys. Lett.} {\bf 214B}, 14 (1988);\hfill
\break
A. Albrecht and N. Turok, {\it Phys. Rev.} {\bf D40}, 973 (1989);\hfill \break
F. Bouchet and D. Bennett, {\it Phys. Rev.} {\bf D41}, 720 (1990).
\item{44.~} G. Smoot et al., Goddard preprint, (1992).
\item{45.~} D. Bennett, A. Stebbins and F. Bouchet, Fermilab preprint,
(1992).

\np

\bigskip
\centerline{\bf FIGURES}

\noindent {\bf Figure 1.} The redshift, $z_{CO}$, at which
photons with energy $E$ are cut off by pair-production
off cosmic background photons. $z=1$ in the present epoch.

\noindent {\bf Figures 2 (i) - (vi).} The photon background from cusp
annihilation as a function of $E$ for
$G\mu/c^2 =10^{-11} - 10^{-16}$ respectively.
$\gamma = 100$ and $\nu=0.03$.
Slicing the plot at $E=10^5$ GeV, the curves
represent $Q_f=10^{18}$ GeV, $10^{14}$ GeV,
$10^{10}$ GeV and $10^{6}$ GeV in order of increasing flux.
The dashed triangle up to $10^6$ GeV is an extrapolation
of the $35-150$ MeV extragalactic photon background.
The UHE data points are the cosmic ray measurements from
the Fly's Eye detector which represent an upper limit on the photon flux at
those energies.

\noindent {\bf Figure 3.} The predicted photon flux as a function of
$E$ for $G\mu/c^2 =10^{-15}$ and various values of $Q_f$, the mass of the
initial superheavy particles emitted
from the cusp. $\gamma = 100$ and $\nu=0.03$
The dashed triangle is an extrapolation
of the $35-150$ MeV extragalactic photon background.
The UHE data points are the upper limit on the photon flux from
the Fly's Eye detector. Slicing the plot at $E=10^5$ GeV, the curves
represent $Q_f=10^{18}$ GeV, $10^{14}$ GeV,$10^{13}$ GeV,$10^{12}$ GeV,
$10^{11}$ GeV, $10^{10}$ GeV and $10^{6}$ GeV in order of increasing flux.

\noindent {\bf Figure 4.} The predicted photon flux as a function of
$E$ for various values of $Q_f=\sigma$ where $\sigma = (\mu\hbar^3)^{1/2}$.
$\gamma = 100$ and $\nu=0.03$. The dashed triangle is an extrapolation
of the $35-150$ MeV extragalactic photon background.
The UHE data points are the upper limit on the photon flux from
the Fly's Eye detector. Slicing the plot at $E=10^8$ GeV, the curves
represent $G\mu/c^2=10^{-11}$, $G\mu/c^2=10^{-16}$, $G\mu/c^2=10^{-12}$,
$G\mu/c^2=10^{-13}$, $G\mu/c^2=10^{-14}$ and $G\mu/c^2=10^{-15}$
in order of increasing flux.

\end

\item{.~} , {\it .} {\bf }, (19).

\end